\documentclass[useAMS,usenatbib]{mn2e}

\usepackage{amsmath, amssymb}
\usepackage[]{graphicx}
\usepackage{epsfig}

\usepackage{lscape}

\newcommand{\rlight}{r_{\rm L}}
\newcommand{\aap}{A\&A}
\newcommand{\mnras}{MNRAS}

\newcommand{\apj}{ApJ}
\newcommand{\apjl}{ApJL}

\newcommand{\nat}{Nature}

\newcommand{\pof}{Physics of Fluids}

\def\ltsima{$\; \buildrel < \over \sim \;$}
\def\simlt{\lower.5ex\hbox{\ltsima}}

\def\gtsima{$\; \buildrel > \over \sim \;$}
\def\simgt{\lower.5ex\hbox{\gtsima}}

\title{Explosive reconnection of double tearing modes in relativistic plasmas: application to the Crab flares}

\author[H. Baty, J. Petri, and S. Zenitani]{H. Baty$^{1}$\thanks{E-mail: hubert.baty@unistra.fr} , J.
Petri$^{1}$\footnotemark[1]
\thanks{E-mail: jerome.petri@unistra.fr} , and S. Zenitani$^{2}$\thanks{E-mail:
seiji.zenitani@nao.ac.jp} \\
$^{1}$Observatoire Astronomique de Strasbourg, 11 Rue de l'Universit\'e, 67300 Strasbourg, France\\$^{2}$National
Astronomical Observatory of Japan, 2-21-1 Osawa, Mitaka, Tokyo 181-8588, Japan\\
}

\begin{document}

\date{Accepted ........... Received ..........; in original form ..........}

\pagerange{\pageref{firstpage}--\pageref{lastpage}} 
\pubyear{2013}

\maketitle

\label{firstpage}

\begin{abstract}
Magnetic reconnection associated to the double tearing mode (DTM) is investigated by means of resistive relativistic magnetohydrodynamic (RRMHD) simulations. A linearly unstable double current sheet system in two dimensional cartesian geometry is considered. For initial perturbations of large enough longitudinal wavelengths, a fast reconnection event is triggered by a secondary instability that is structurally driven by the nonlinear evolution of the magnetic islands. The latter reconnection phase and time scale appear to weakly depend on the plasma resistivity and magnetization parameter. We discuss the possible role of such explosive reconnection dynamics to explain the MeV flares observed in the Crab pulsar nebula. Indeed the time scale and the critical minimum wavelength give constraints on the Lorentz factor of the striped wind and on the location of the emission region respectively.
\end{abstract}

\begin{keywords}
plasmas - magnetohydrodynamics - magnetic reconnection - relativity
\end{keywords}

\section{Introduction}

Magnetic reconnection is a key process for the dissipation of magnetic energy in a variety of astrophysical and laboratory plasmas. This is particularly the case for pulsars and magnetars environments, which manifest strong flares in X-rays and gamma-rays \citep{2005Natur.434.1107P, 2011ApJ...741L...5S}. This bursty activity is usually attributed to the sudden release and conversion of magnetic energy into other forms of energy, let it be thermal and/or kinetic, in a way similar to solar flares \citep{2000mare.book.....P}. 

The presence of current sheets around strongly magnetized neutron stars suggest that the tearing instability may be a good candidate to drive reconnection \citep{2003MNRAS.346..540L}. The linear and nonlinear development of the single tearing mode in relativistic magnetically dominated regime has been investigated in the literature (\cite{2007MNRAS.374..415K} and references therein). Their results show the release of magnetic energy via the formation and growth of magnetic islands on a Sweet-Parker-like time scale proportional to $S^{1/2}$, $S$ being the Lundquist number defined by $S = l c/\eta$, where $l$ is the current sheet thickness, $c$ the light speed, and $\eta$ the resistivity. This scaling follows the same law as in the non relativistic regime \citep{2000mare.book.....P}. In this work, the Lundquist number~$S$ is considered to be $S\gg1$.

As a consequence, it is  very uncertain that the relativistic tearing mode can grow on a time-scale 
fast enough to explain the short flare-rising time observed in pulsar wind nebula like in the Crab Nebula ($\sim$ a few hours, see \cite{2011ApJ...741L...5S}).
Moreover, the plasma flow associated to the tearing instability is largely sub-Alfv\'enic and sub-relativistic, with a value that is at least  two orders of magnitude smaller than the value deduced from the relativistic bulk flow expected from such energetic flares \citep{2012ApJ...749...26B}. Consequently,
extremely short time scales deduced from such highly energetic  flares is appealing and probably requires another mechanism. 

The magnetic structure of the pulsar magnetosphere involves the presence of a current sheet wobbling around the equatorial plane \citep{1990ApJ...349..538C}. It is well known that such configurations are subject to multiple tearing modes, rather well studied in laboratory magnetic confinement systems (see for example \cite{2005PhPl...12h2504B} and references therein). A particularly interesting result has been obtained in  the non relativistic framework, where a fast reconnection phase has been identified during the (slower) nonlinear evolution  of double tearing modes (DTM) \citep{1980PhFl...23.1368P}. It manifests itself as a secondary instability leading to the merging of the magnetic islands initially situated on the two separated current layers \citep{2002PhRvL..89t5002I, 2011PhRvL.107s5001J}. Interestingly, the process is considered to be explosive as the time scale of the instability is shown to weakly depend on the Lundquist number, scaling as $\sim S^{\alpha}$ with $\alpha$  in the range [0. : 0.2] \citep{2007PhRvL..99r5004W}. The associated plasma flow is also shown to be much stronger than the value obtained from a single tearing mode evolution, with a magnitude approaching the Alfv\'en speed.

The aim of the present study is precisely to extend these results to relativistic plasmas. One must note that, mainly for numerical reasons, we limit ourself to only mildly relativistic velocities with an Alfv\'en speed~$c_A \simlt 0.8 c$. We use RRMHD simulations in a simple two-dimensional cartesian geometry. The nonlinear dynamics is considered with respect to varying longitudinal wavelengths~$\lambda$, changes in the Lundquist number $S$ and in the magnetization parameter~$\sigma$. The paper is organized as follows. The model and numerical setup are described in \S~\ref{sec:Modele}. In \S~\ref{sec:Results}, the results are presented. Finally, a discussion on the consequences for pulsar wind nebula and neutron star magnetospheres is considered in \S~\ref{sec:Discussion}, and a conclusion is drawn in \S~\ref{sec:Conclusion}.

\begin{figure}
\centering
\includegraphics [scale=0.19]{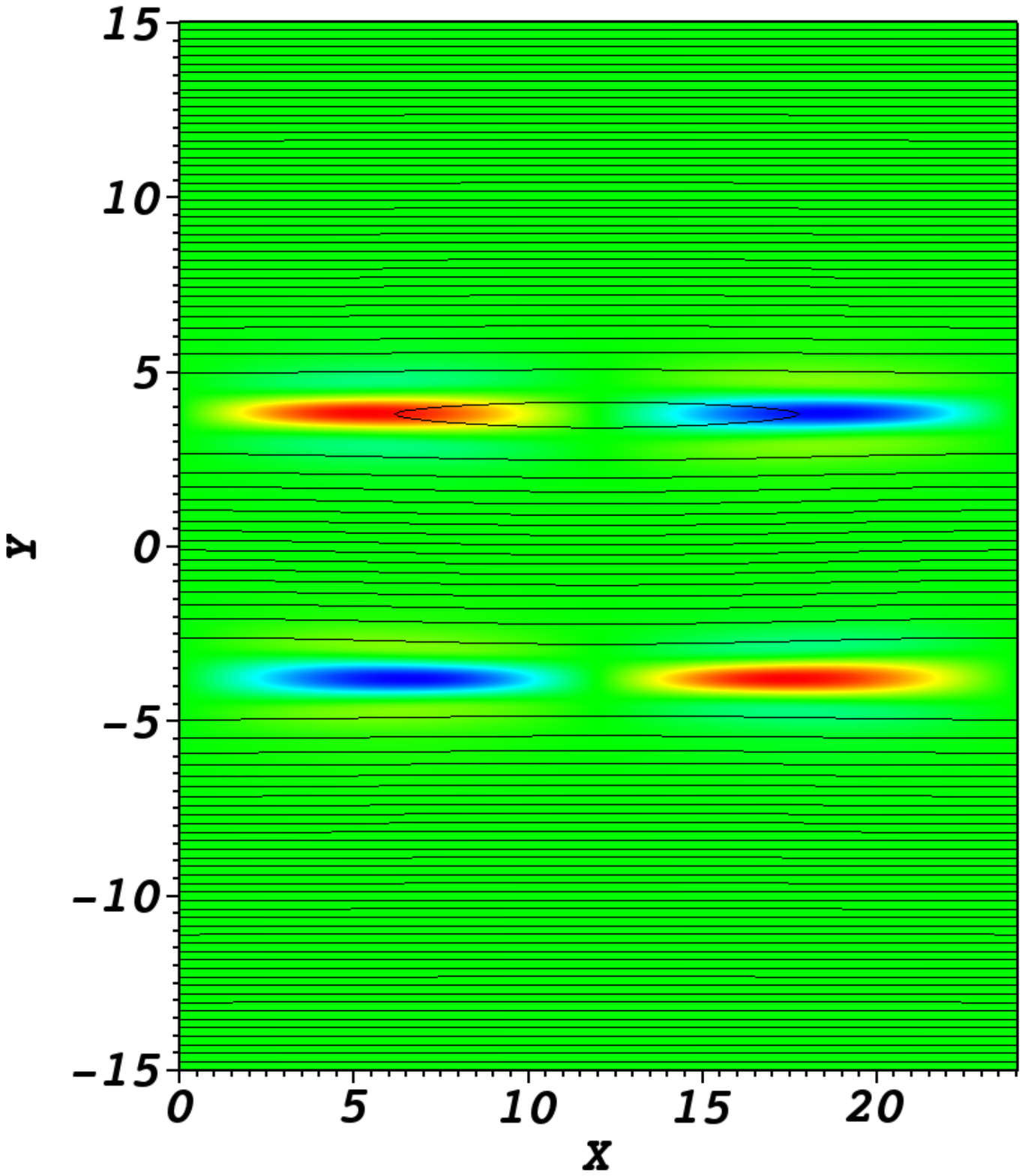}
\includegraphics [scale=0.19]{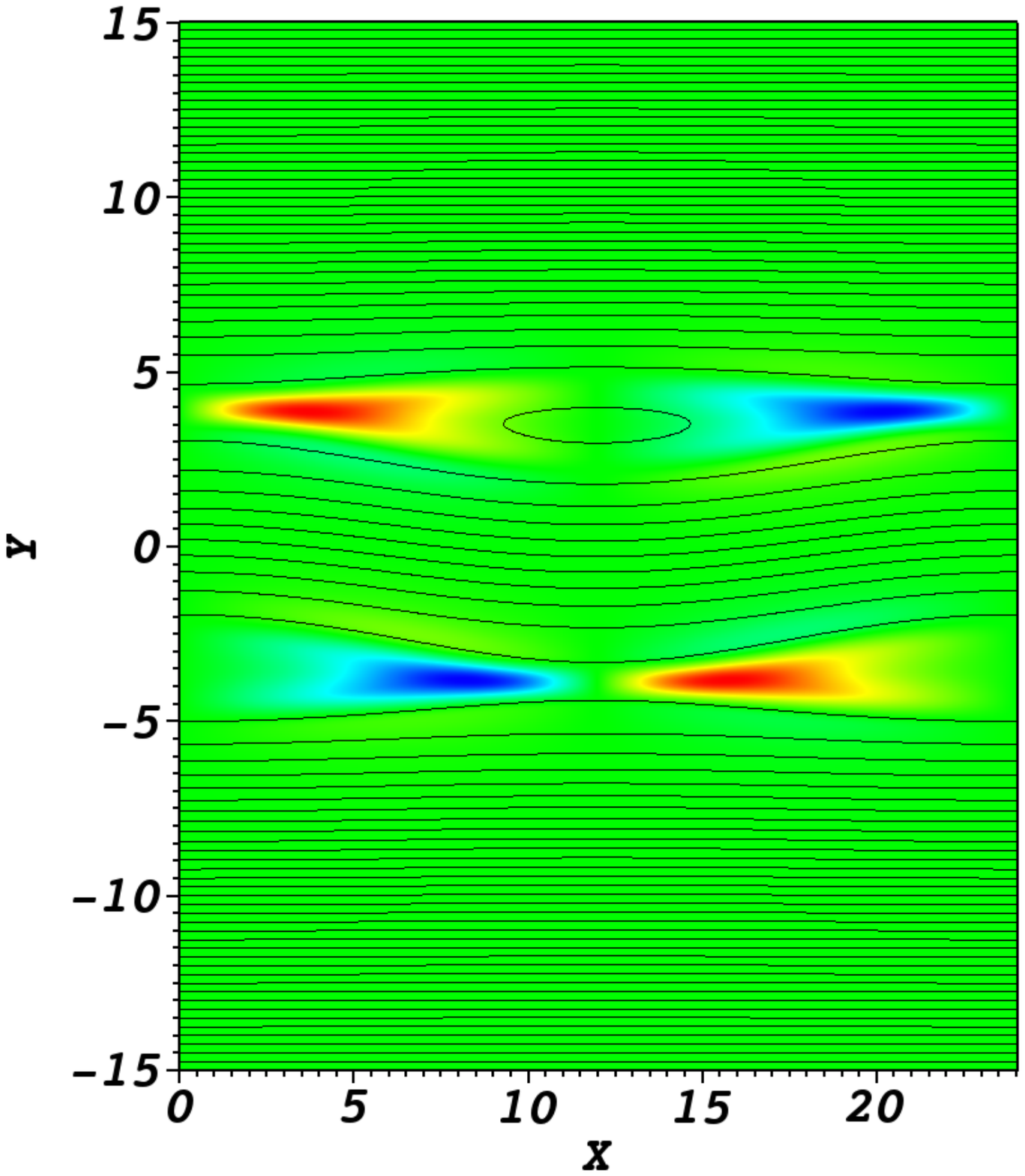}
\includegraphics [scale=0.19]{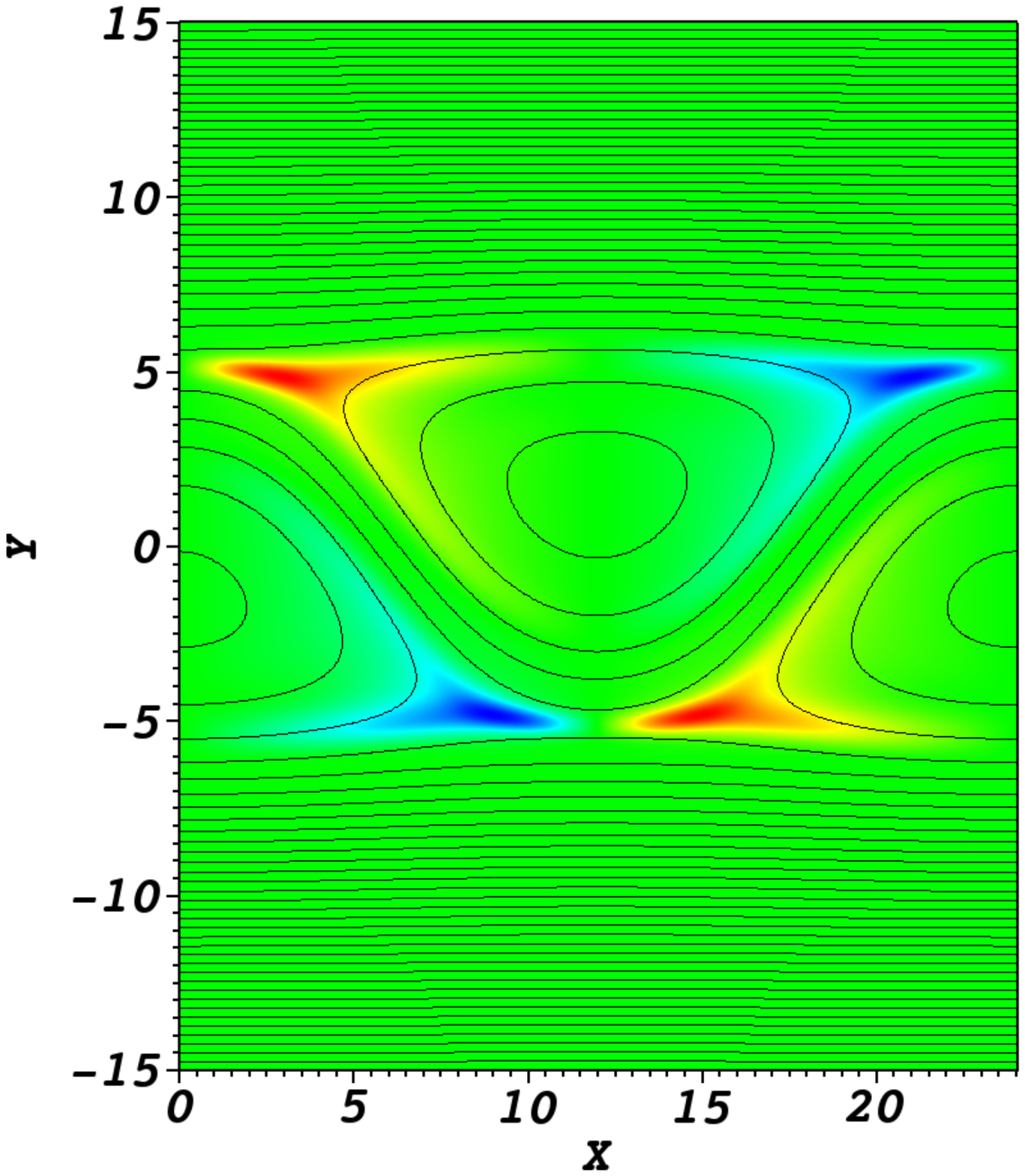}
\includegraphics [scale=0.19]{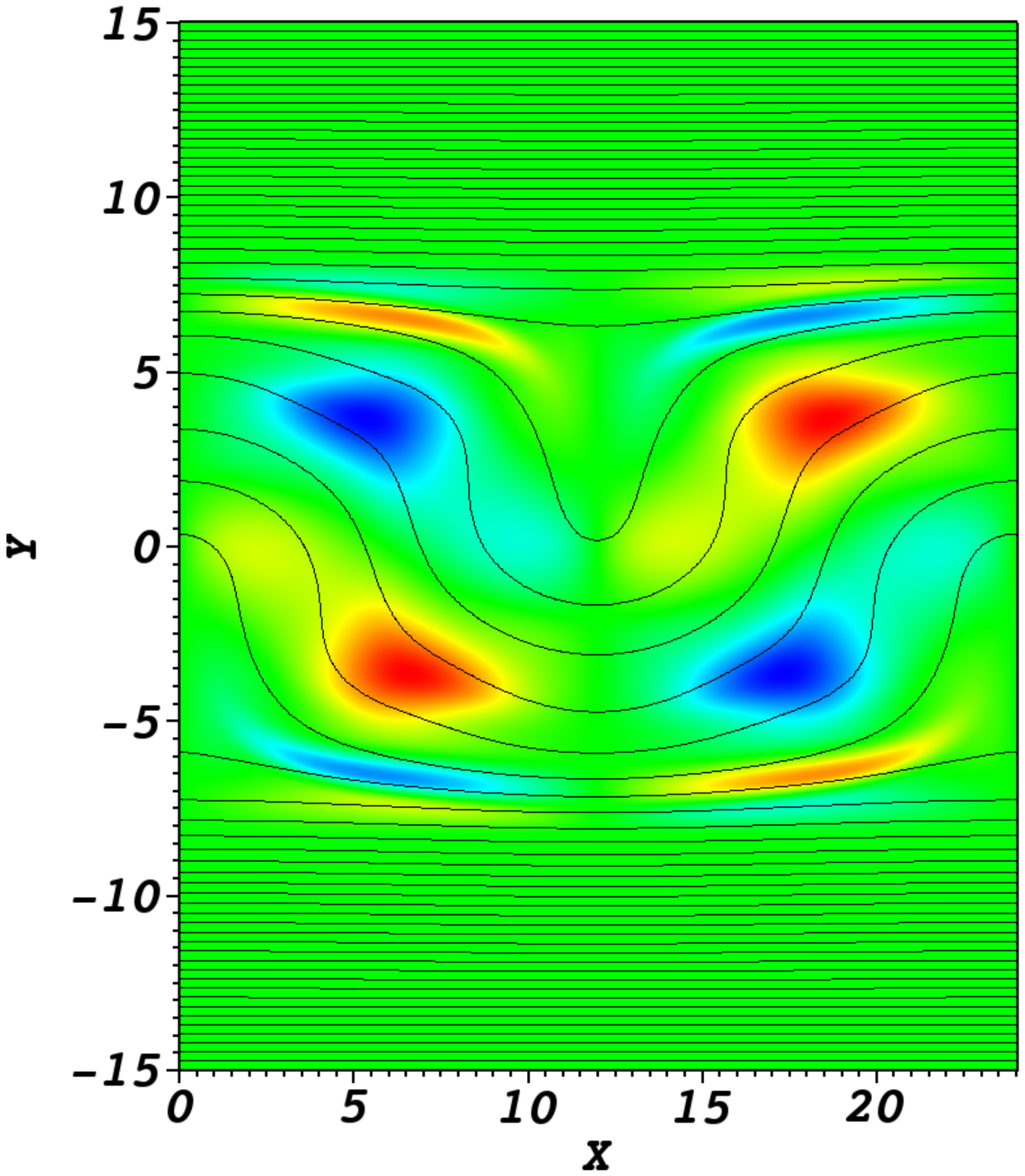}
\caption{$V_x$ component of the flow velocity overlaid with magnetic
field lines at four different times, from left to right and up to down: $t = 400 (a), t = 1000 (b),
t = 1500 (c), t = 1800 (d)$. The simulation is using $S = 200$, $L_x = 24$, and
$\rho_g = 1$. The contour levels normalization for the velocity varies
from panel to panel (negative values are in blue and positive values in red).}
\label{fig1}
\end{figure}

\section[]{Model and numerical setup}
\label{sec:Modele}

We perform numerical simulations by using the RRMHD equations described in detail by \cite{2010ApJ...716L.214Z}. These equations are solved by a variant of \cite{2007MNRAS.382..995K} time-split HLL method, where the stiff non-ideal terms are treated using a relaxation scheme. A hyperbolic divergence cleaning method is also used in order to deal with the solenoidal condition \citep{2002JCoPh.175..645D}. For more details about the code, the reader is referred to \cite{2010ApJ...716L.214Z} and references therein. The light speed is conveniently normalized to unity, $c = 1$. We start with an initial double Harris-sheet-like configuration
\begin{equation}
  \label{eq:Bprofile}  
  \vec{B} = B_0  \left[  1 + \tanh \left[ (y - y_0)/l \right] - \tanh \left[ (y + y_0)/l \right]  \right]  \vec {x},       
\end{equation}
\noindent
where $B_0$ is the maximum magnetic field amplitude. The initial temperature~$T$ is assumed to be uniform, and the plasma density variation follows the expression
\begin{equation}
  \label{eq:Rprofile}  
              \rho =  \rho_b + \rho_0  \left[  \cosh ^{-2} \left[  \left( y  - y_0 \right) /l \right] + \cosh ^{-2} \left[ (y  + y_0)/l \right] \right].
\end{equation}
\noindent
This corresponds to a structure in equilibrium as the thermal pressure is taken to be $P = \rho$. The amplitude of the magnetic field is set so that $B_0^2 / 2 = 1$. Throughout our study, the current sheet thickness is taken to be $l = 1$ and the half-separation between the two current layers is $y_0 = 4$. We also fix the specific heat ratio at~$\gamma=4/3$. The initial density values are normalized, by using $\rho_0 =1$, whilst the background value $\rho_b$ varies from case to case. Typically, we choose 4 values for this background density, namely $\rho_b = 2, 1, 2/3$ and $1/3$, leading to 4 values for the corresponding background Alfv\'en speed such that $c_A = 0.447 , 0.535, 0.612,$ and  $0.739$. The Alfv\'en speed is defined by $c_A^2 =  B_0^2/( \rho  h + B_0^2)$, where $h$ is the specific enthalpy, $h =  1 + \gamma P /  \left[ \rho(\gamma - 1)\right]$.
Note that, the background density is also a measure of the magnetization of the system, as the
magnetization parameter $\sigma$ follows the relation: $c_A =  \left[ \sigma/(\sigma + 1) \right]^{1/2}$ \citep{2007A&A...473...11D}. Consequently, the magnetization parameter values considered in our study are, $\sigma = 0.2, 0.4, 0.6,$ and $1.2$.This magnetization parameter is different from the definition used in \cite{2011ApJ...739L..53T} but similar to the one taken in \cite{2010ApJ...716L.214Z}.

The simulation domain is bounded by a rectangular box of dimensions $[0: L_x]  \times [-L_y /2 : L_y/2]$, with a fixed size value in the perpendicular~$y$ direction $L_y = 30$. The longitudinal size $L_x$ can be adjusted in order to select different wavelengths, as periodic boundary conditions are applied at $x = 0, L_x$. Free boundary conditions are additionally imposed at $y =  \pm L_y/2$. In most of the present study, a uniform resolution of $20$ grid points per unit length scale is chosen (i.e. $600$ points across the full box width $L_y = 30$). We have checked that the latter grid resolution allows us to carry out calculations with a maximum Lundquist number of order $500$, lower resistivity (e.g. higher Lundquist numbers) simulations being dominated by the numerical resistivity due to truncation errors inherent to the numerical scheme. Note also that a similar resolution is used in \cite{2010ApJ...716L.214Z}.

A small perturbation with a maximum magnitude $ \simeq 10 ^{-4} \times B_0$ is added at $t = 0$ to the magnetic field components in order to trigger a tearing perturbation on each layer situated at $y = \pm y_0$.

\begin{figure}
\includegraphics [width=0.5\textwidth]{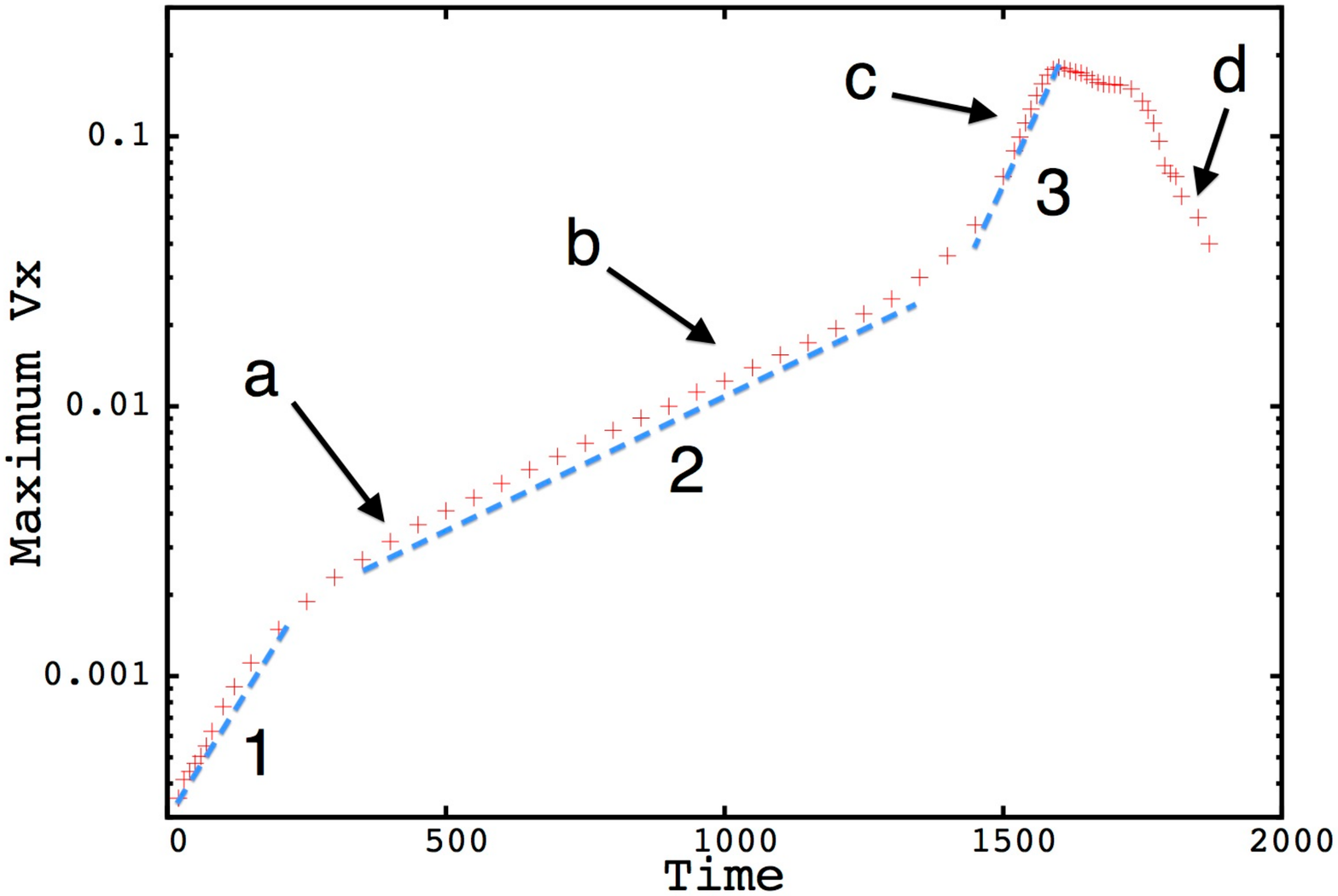}
\caption{Maximum $V_x$ flow velocity as a function of time for the simulation
of Figure 1. The four corresponding panels of Figure 1 are indicated by the label a, b, c, and d respectively.
Three stages corresponding to three distinct regimes (see text) are indicated
by adding a broken line.}
\label{fig2}
\end{figure}

\begin{figure}
\includegraphics [width=0.5\textwidth]{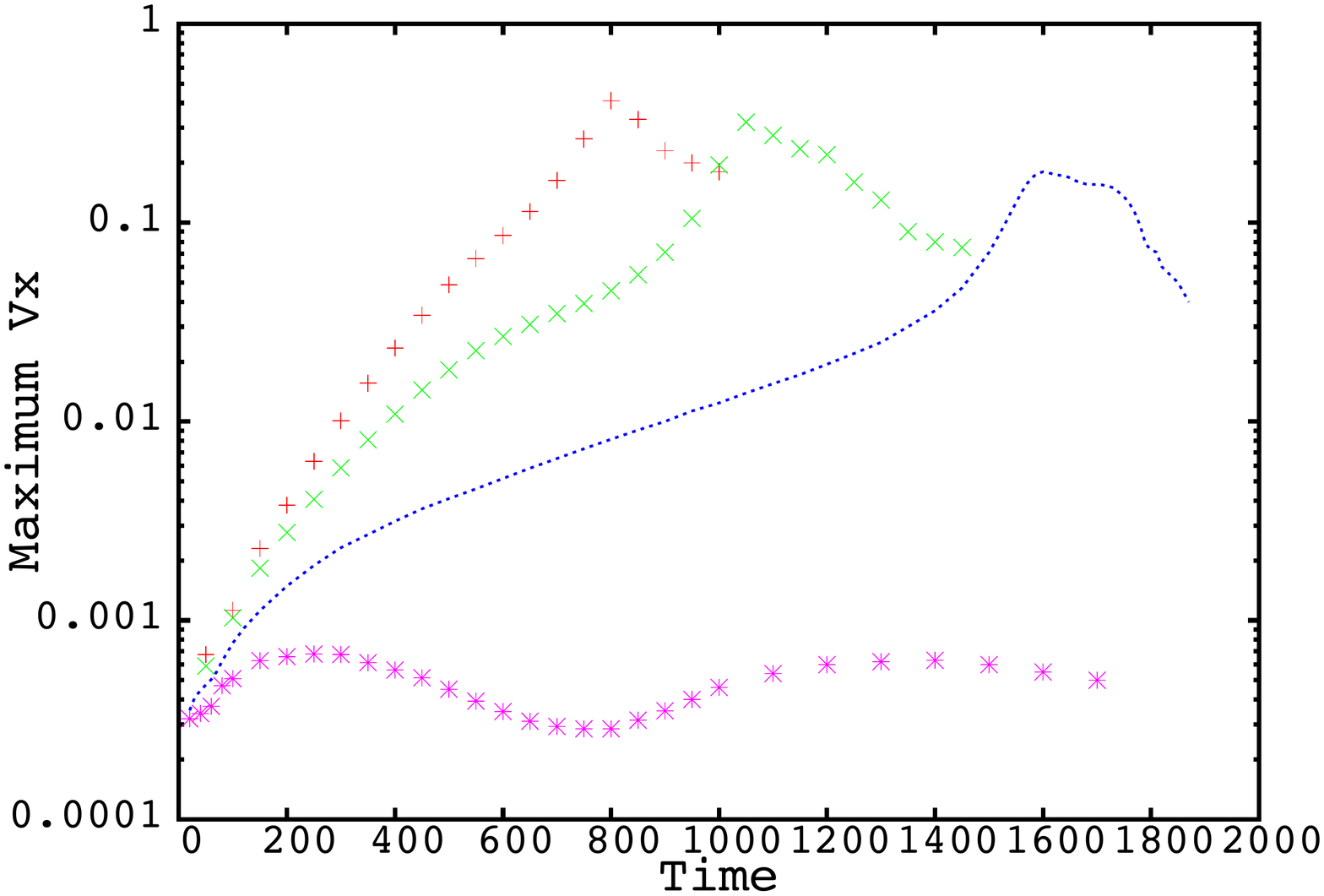}
\caption{Maximum $V_x$ flow velocity as a function of time for runs selecting four different
longitudinal lengths: $L_x = 40, 30, 24$, and $20$ (from uppermost to downermost  curves).
We use  $\rho_b = 1$, and
$S = 200$. }
\label{fig3}
\end{figure}

\section{Results}
\label{sec:Results}

The time evolution of a typical run with $S=200$, a longitudinal length set to~$L_x  = 24$, and with $\rho_b = 1$, is illustrated in Figure~\ref{fig1} and Figure~\ref{fig2}. For simplicity, we have arbitrarily chosen the spatial $x$ component of the four-velocity, $V_x$, as a diagnostic measure of the instability magnitude. 
In Figure~\ref{fig2}, it can be seen that the system evolution exhibits four distinct phases, three growing phases indicated by 3 hatched lines, and a last fourth decaying $V_x$ phase. The different phases can be connected to the corresponding plates of Figure~\ref{fig1}, showing $V_x$ overlaid with the magnetic field lines. Phase 1 (first pannel of  Figure~\ref{fig1}) corresponds to the early growth of two well separated magnetic islands situated at the two current layers at $L_y = \pm 4$. Note that the upper island (with the $O$-point at $L_x = 12$) is shifted by one half longitudinal wavelength compared to the lower island (with the $O$-point located at $L_x = 0, 24$. It corresponds to the fact that an asymmetric instability is more unstable than a symmetric one. The time scale for this first phase roughly corresponds to a Sweet-Parker-like process $\sim S^{1/2}$ (in agreement with the non relativistic results, \cite{2007PhRvL..99r5004W}). A transition to a second phase (2) then occurs, but with islands continuing to grow (second pannel of Figure~\ref{fig1}) on a somewhat slower time scale. Once a threshold is reached, visible at $t \ge 1400$, the two islands start to grow in an more explosive way (Phase 3). During this phase, the islands interact more and more leading to a triangular-shape deformation (third pannel of Figure~\ref{fig1}). As a feedback, this structural deformation is able to drive a magnetic reconnection characterized by a merging process between the two islands at a time around the peak observed at $t = 1600$ in Figure~\ref{fig2}. Finally, a last phase is observed, where $V_x$ decays and the magnetic configuration relaxes toward a new stable state (last pannel of Figure~\ref{fig1}). The final state is observed to be free from magnetic islands and current layers, and the corresponding magnetic field lines tend to become straight again. We have checked that, during the whole process, the initial magnetic energy situated in the central region has been transformed into kinetic bulk flow and also into thermal energy, in roughly equal proportion.

The previous simulation shows a behaviour similar to what is expected from non relativistic simulations \citep{2011PhRvL.107s5001J, 2011PhPl...18j2112J, 2011PhPl...18e2303Z}. We have first checked the dependence with the longitudinal wavelength $L_x$ (see Figure\ref{fig3}). Our results show that the secondary instability is present only when a minimum wavelength is selected, with a threshold value between $20$ and $24$ in our case. Otherwise, the growth of the two magnetic islands is observed to saturate with the perturbed velocity relaxing to zero (see the lower curve corresponding to a run using $L_x = 20$ in Figure 3). This roughly agrees with the results of \cite{2011PhRvL.107s5001J, 2011PhPl...18j2112J} for which a criterion $L_x / y_0   \simgt  6$ was observed to be necessary for triggering the explosive secondary growth. Moreover, the run with $L_x = 20$ gives an estimate of the maximum velocity flow associated to a single tearing mode, thus confirming that the speed is very sub-Alfv\'enic and sub-relativistic (i.e. $2-3$ orders of magnitude lower compared to the runs where the secondary explosive growth is triggered).

\begin{figure}
\includegraphics [width=0.45\textwidth]{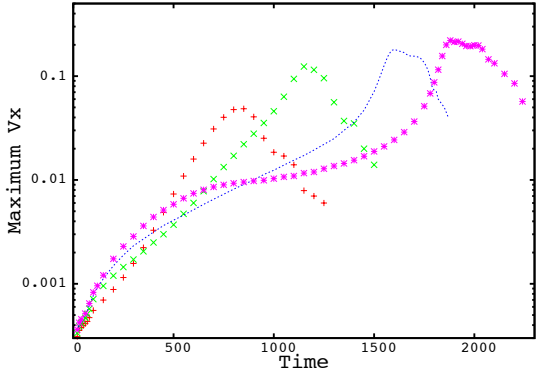}
\caption{Maximum $V_x$ flow velocity as a function of time for runs using four different resistivity
values: $S = 50, 100, 200$, and $400$ (from leftermost to rightermost curves) . We use $L_x = 24$, and
$\rho_b = 1$.}
\label{fig4}
\end{figure}

Second, we have examined the influence of the Lundquist number~$S$. The results are plotted in Figure~\ref{fig4} for a simulation using $L_x = 24$ and $ \rho_b = 1$, for four different values of $S$ (in the range $50-400$). A close inspection of the time scale involved during the explosive growth (Phase 3) demonstrates that it only weakly depends on the resistivity. Even, mainly for numerical requirements, we were limited to a small range of $S$ values in this study, our estimate gives a tendency for a scaling close to $S^{0.2-0.3} $. The same conclusion holds if one takes the observed maximum $V_x$ velocity. A different scaling behavior is however evident (in Figure 4) when inspecting the effect of $S$ on the second phase. Indeed, we have found that the growth of this latter phase can be fitted with a scaling close
to a purely resistive dependence $ \sim S$. This result agrees with the nonlinear evolution of a single
magnetic island entering the saturation-Rutherford regime \citep{2007PhRvL..99r5004W}.

\begin{figure}
\includegraphics [width=0.5\textwidth]{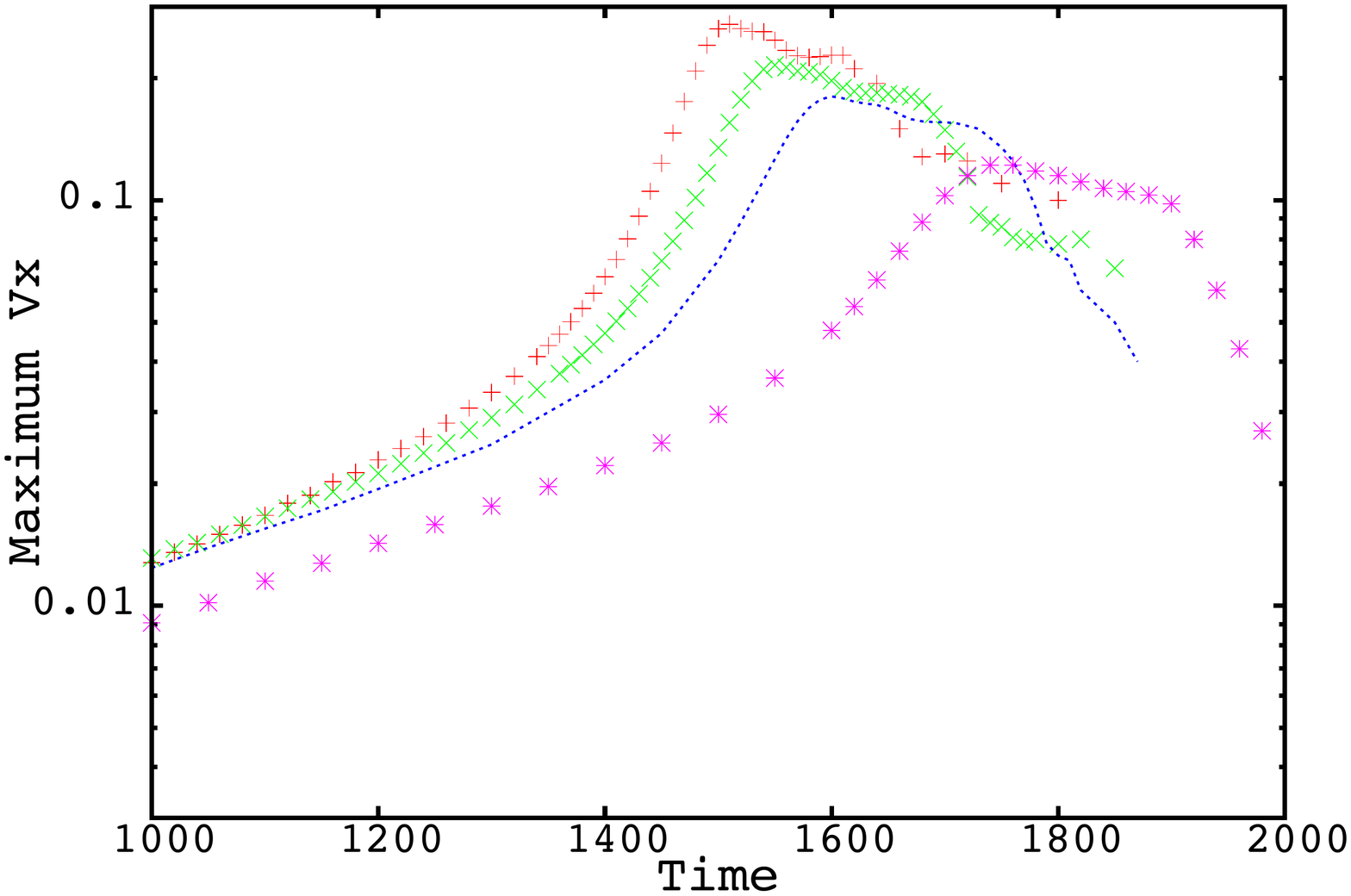}
\caption{Maximum $V_x$ flow velocity as a function of time for runs using four different
background density values: $\rho_b = 1/3, 2/3, 1,$ and $2$ (from leftermost to rightermost curves). We use $L_x = 24$, and
$S = 200$. Note that only the final stages are shown.}
\label{fig5}
\end{figure}

Finally, we have examined the effect of varying the background density $\rho_b$, in order to investigate the effect of the magnetization parameter. The results are plotted in Figure~\ref{fig5} for a somewhat moderate range of four different values. Again,  relatively weak dependence following $\sigma^{0.3} $ can be deduced from our simulations by using the measured time scale or the maximum velocity as well.

As concerns the maximum velocity obtained during the simulations, the data plotted in the previous figures clearly show that outflow relativistic speeds of order $\sim 0.5c$ can be easily  achieved in our mildly magnetized configurations. We can therefore expect that configurations having higher magnetization parameter and also higher Lundquist number should exhibit even faster flows.

\section{Discussion}
\label{sec:Discussion}

The explosive character of the reconnection process is typical of a double current sheet layer. Those magnetic configurations exist in the so called striped wind of a pulsar. It is therefore possible that this mechanism releases quickly a substantial fraction of the energy flux entailed in the Poynting dominated flow. An observational manifestation would be the gamma-ray flares recently discovered in the Crab nebula 
\citep{2012ApJ...749...26B, 2013ApJ...765...52S}. Those flares came as a real surprise. It was long believed that the Crab nebula was a steady state emitter. Moreover the variability on time scales of a few hours to a few days is too long to be immediatly connected to the pulsar activity with period of 33~ms. It is also too short to be attributed to the secular evolution of the nebula itself. \cite{2012MNRAS.427.1497L} argued that a kink instability in the polar region of the wind is responsible for the gamma-ray emission.

In this work we claim that 
localized reconnection in the unschocked part of the striped wind triggers explosive reconnection which creates a radially sheared flow producing internal shocks as in the GRB models (see also \cite{2012MNRAS.426.1374C} for another model with reconnection). A substantial fraction of the pulsar spin-down energy could be released by a clump of matter (i.e. plasmoid) moving outwards at relativistic speed and supplemented with fresh particles from the inner part of the wind. 
Magnetic reconnection is known to form plasmoids of hot plasma, which in our case are advected by the relativistic motion of the wind.
Let us discuss in more detail the orders of magnitude. The above simulations showed that the rising time of the explosive phase last about 160 units of time. In the striped wind interpretation, the separation between the current sheets,~$2\,y_0$, is equal to half a wavelength $\lambda_L\approx2\pi\rlight$ where $\rlight=c\,P/2\pi$ is the light-cylinder radius and $P$ the period of the pulsar.
This corresponds to the maximal separation of two current sheets when viewed in the equatorial plane. At higher latitudes, the separation becomes smaller and the explosive reconnection phase could be triggered earlier. To fix orders of magnitude, we choose in this discussion current sheets in the equator. According to Lorentz-time dilation and length contraction between the wind frame and the observer frame, the duration of the explosive rise as seen by this distant observer is $\Delta T \approx 10 \Gamma^2 \, P$ (we took $y_0=4$), $\Gamma$ being the Lorentz factor of the wind. Applied to the Crab with $P=33$~ms with typical rising time of about~$\tau_r\approx10$~hr \citep{2013ApJ...765...52S}, the duration of the explosive time should be less than $\Delta T \lesssim \tau_r$~hr.  We therefore infer a maximum Lorentz factor of $\Gamma\lesssim300$ which is consistent with the magnitude expected at the base of the wind close to the light-cylinder where from independent models~$\Gamma\approx20-50$ \citep{2005ApJ...627L..37P}. Therefore we conclude that explosive reconnection occurs outside the light-cylinder but not necessary very far away from the pulsar. We believe that the flares are still moving relativistically, dragged by the bulk motion of the wind, Doppler beaming and frequency shift are important. The peak energy observed at 380~MeV is thus only 1~MeV in the fluid frame and consistent with the synchrotron uppermost limit at about 250~MeV. Relativistic beaming is favored by the correlation between peak energy and flux as discussed 
in~\cite{2012ApJ...749...26B}. In~\cite{2012MNRAS.424.2023P} we showed that Fermi observations can be reconcilied 
with synchrotron emission in the striped wind including reconnection. The current sheet geometry was supposed stationary leading to pulsed emission. Reconnection destroys the regularity of the magnetic field, we then expected a disappearance of the pulsed emission but still MeV/GeV photons without any periodicity and associated with flares, a radially moving blob disconnected from the striped wind. Because the wind magnetization is high, magnetic energy liberation dominates the spin-down luminosity of the pulsar, increasing the flux received at Earth. A typical energy within a flare of one day~$t_f\approx1$~d is $E_f\approx10^{34}$~J. The typical area of the reconnection layer is a square of length~$\lambda_L$. During this flare, the energy passing through this area is
$E_f \approx B^2 \, \lambda_L^2 \, c \, t_f / \mu_0$. In order to get these energies of~$E_f$ the local magnetic field intensity in the flare has to be around $2$~T. Taking into account the wave nature of the striped wind with $B\approx B_L\,\rlight/r$ and $B_L\approx100$~T, the emission region is located at $r\approx50\rlight$. As a consequence, flares could emanate from the same site as the one producing the pulsed gamma-ray emission, location and Lorentz factor are consistent in both cases. At this stage of our model, it is impossible to discuss about their respective spectra because this requires a kinetic description of the reconnection site which is out of the scope of this study. 
Furthermore, at this stage of investigation of the present work, we are not able to carry out an analysis to explain the rarity of the observed flares (a few events per year)
\citep{2011ApJ...741L...5S}. More research is needed for quantitative prediction, both in DTM theory and in Pulsar Wind Nebula theory.

\section{Conclusion}
\label{sec:Conclusion}
The multiphase nonlinear development of DTM has been studied using RRMHD simulations
of magnetically dominated plasmas. It starts with an early $S^{1/2}$ Sweet-Parker-like phase
characterized by magnetic islands growth, followed by a saturation in agreement with
a slower $S^1$ Rutherford regime. A third phase exhibiting an explosive growth is then triggered when
the longitudinal wavelength exceeds a critical value, that is $\sim 6$ times the half-separation
length between the two current layers of our simple model. As previously obtained in classical MHD, we suggest
that it represents a secondary instability driven by a geometrical deformation of the magnetic
islands \citep{2011PhPl...18j2112J}. It manifests by a fast merging process between the two islands.
Our results also indicate a relatively weak dependence of the time scale of the explosive phase with 
the resistivity and the magnetization parameter as well (i.e.  $\sim S^{0.2-0.3}$ and $\sigma^{0.3} $ respectively). 

Mainly for numerical reasons, in this work, we limit ourself to values of $S$ and $\sigma$ that are probably
only moderately high compared to more realistic expected values in pulsars environments.
More numerical simulations are therefore needed in order to explore higher values. The recent MHD codes developed by \cite{2011MNRAS.418.1004Z}  
and \cite{2009MNRAS.394.1727P} seem to be very promising tools in this respect.

In this study, we also focused on the DTM instability using a very simple configuration. As pulsar winds however exhibit
more complex multiple current layers, multiple (or at least triple) tearing modes should be also explored.

Anyway, DTM is a better candidate than the simple tearing mode in order to explain the energetic Crab flares in the context of magnetic reconnection. Indeed, the explosive growth phase clearly involves a faster time scale
accompanied by a relativistic bulk flow that is not present in a simple tearing mode. Moreover, it puts some interesting constraints (see the above discussion)
on the maximum Lorentz factor of the striped wind {\bf  ($\Gamma\lesssim300$) } and on the localization of the emission region respectively ($r\approx50\rlight$).

\bibliographystyle{mn2e}

\end{document}